\documentstyle[amssymb,12pt,thmsa,sw20lart]{article}

\input{tcilatex}
\begin{document}

\title{The classical-statistical limit of quantum mechanics.}
\author{Mario Castagnino \\
Instituto de Astronom\'{i}a y F\'{i}sica del Espacio. Casilla de Correos 67,%
\\
Sucursal 28, 1428 Buenos Aires, Argentina.}
\maketitle

\begin{abstract}
The classical-statistical limit of quantum mechanics is studied. It is
proved that the limit $\hbar \rightarrow 0$ is the good limit for the
operators algebra but it si not so for the state compact set. In the last
case decoherence must be invoked to obtain the classical-statistical limit.
\end{abstract}

\section{Introduction}

In this talk, using the methods of papers \cite{Pra}, \cite{Pre}, and \cite
{Pl}, we will consider the classical-statistical limit of quantum mechanics
for a system with continuous evolution spectrum (We have discussed how
classical-statistical limit of quantum mechanics becomes the proper
classical limit, i. e. the localization phenomenon, in paper \cite{Deco}).

This study seems necessary because the existent einselection-approach theory
is far from being perfect since, at least, the following features must be
better explained:

i.- Einselection is based in the decomposition of the system in both a
relevant part, the{\it \ proper system,} and an irrelevant one: the {\it %
environment}. This decomposition it not always possible, e. g. in the case
of the universe, in fact:

{\it ''...the Universe as a whole is still a single entity with no 'outside'
environment, and therefore, any resolution involving its division in
unacceptable'' (}Zurek{\it \ }\cite{1994}{\it , }p. 181{\it ).}

ii.- Moreover there is not a clear criterion to define the ''cut'' between
the proper system and its environment. In fact:

{\it ''In particular, one issue which has been often taken for granted is
looming big, as a foundation of the whole decoherence problem. It is the
question of what are the 'systems' which play a such crucial role in the
discussions of the emergence classicality. This issue was raised earlier,
but the progress to date has been slow at best''(}Zurek{\it \ }\cite{1998}%
{\it , }p. 122{\it ).}

iii.- The definition of the basis where the system becomes classical, i.e. 
{\it the pointer basis,} is based in a predictability sieve which would
produce the set of most stable possible states. But in practice this
definition seems very difficult to implement. In fact, the basis vectors o
are {\it jut good candidates for reasonable stable} states \cite{2003}.

None of these problems seem solved by the einselection approach today \cite
{2003}. Nevertheless when in particular models the three mentioned problems
can be bypassed the einselection approach works extraordinary well.

On the other hand in the ''self-induced'' approach that we are about to
explain all these problems are absent: it can be used in close systems as
the universe\cite{Univ.}, the definition of a convenient subalgebra takes
the role of the coarse-graining induced by the environment \cite{Olimpia},
and the pointer basis for the classical limit is perfectly defined (we will
not consider this problem here since it is fully explained in \cite{Deco}).
In this way the new approach is justified.

\section{The formalism for the observables.}

Let us begin defining the quantum and classical operator algebra and their
limits. We will consider a system with a complete set of commuting
observables (CSCO) $\{\widehat{H,}\widehat{P_{1}},...,\widehat{P}_{N-1}\}$,
where $\widehat{H}$ has a continuous spectrum $0\leq \omega <\infty $ and
also the $\widehat{P_{i}}$ have continuous spectra. Since the energy
continuous spectrum is essential for our treatment we present the simplest
model with just the continuous spectrum (we have added discrete spectra in
ref. \cite{Deco} and \cite{Rolo}) To simplify the notation we will just call 
$\{\widehat{H},\widehat{P}\}$ the set $\{\widehat{H,}\widehat{P_{1}},...,%
\widehat{P}_{N-1}\}$. We will consider the orthonormal eigen basis $%
\{|\omega ,p\rangle \}$ of $\{\widehat{H,}\widehat{P}\}$ and write the
hamiltonian and $\widehat{P}$ as 
\begin{equation}
\widehat{H}=\int_{p}\int_{0}^{\infty }\omega |\omega ,p\rangle \langle
\omega ,p|d\omega dp^{N-1}\qquad \widehat{P}=\int_{p}\int_{0}^{\infty
}p|\omega ,p\rangle \langle \omega ,p|d\omega dp^{N-1}  \label{2.1}
\end{equation}
Furthermore we will consider the algebra $\widehat{{\cal A}}$ of the
operator (as we have said we are using the ``final pointer basis'' of paper 
\cite{Deco}): 
\[
\widehat{O}=\int_{p}\int_{0}^{\infty }O(\omega ,p)|\omega ,p\rangle \langle
\omega ,p|d\omega dp^{N-1} 
\]
\begin{equation}
+\int_{p}\int_{p^{\prime }}\int_{0}^{\infty }\int_{0}^{\infty }O(\omega
,\omega ^{\prime },p,p^{\prime })|\omega ,p\rangle \langle \omega ^{\prime
},p^{\prime }|d\omega d\omega ^{\prime }dp^{N-1}dp^{\prime N-1}  \label{2.2}
\end{equation}
where the first term in the r.h.s. will be called $\widehat{O_{S}}$ or the 
{\it singular component} and the second term will be called $\widehat{O_{R}}$
or the {\it regular component}. Functions $O(\omega ,p)$ and $O(\omega
,\omega ^{\prime },p,p^{\prime })$ are regular (see \cite{Pra} for details), 
$[\widehat{H},\widehat{O_{S}}]=0$, $\widehat{O_{S}}\in \widehat{{\cal L}}%
_{S} $, the singular space, $\widehat{O_{R}}\in \widehat{{\cal L}}_{R}$, the
regular space, and $\widehat{{\cal A}}=\widehat{{\cal L}}_{S}\oplus \widehat{%
{\cal L}}_{R}$. The observables are the self adjoint operators of $\widehat{%
{\cal A}}$. As $O(\omega ,\omega ^{\prime },p,p^{\prime })$ is a regular
function $\widehat{{\cal L}}_{R}$ can be promoted to a space of operators on
a Hilbert space ${\cal H}$ and the usual theory of the Wigner function \cite
{Wigner} can be implemented in $\widehat{{\cal L}}_{R}$.

At the classical statistical level $\widehat{{\cal L}}_{R}$ will have a
classical analogue ${\cal L}_{R}$ a space of integrable $L_{1}$ -- functions
over the phase space ${\cal M=M}_{2N}\equiv {\Bbb R}^{2N}$. These functions
will be called $O_{R}(\phi )$ where $\phi $ symbolizes the coordinates over $%
{\cal M}$, $\phi ^{a}=(q^{1},...,q^{N},p^{1},...,p^{N})$ with $a=1,2,...2N$.
As it is known we can map $\widehat{{\cal L}}_{R}$ on ${\cal L}_{R}$ via the
usual Wigner transformation or symbol as $symb:\widehat{{\cal L}}%
_{R}\rightarrow {\cal L}_{R}$ 
\begin{equation}
symb\widehat{f_{R}}=f_{R}(\phi )=\int_{-\infty }^{\infty }\langle {\bf q-y|}%
\widehat{f}{\bf |q+y\rangle }e^{2i{\bf q.y}/\hbar }d^{N}y\ .  \label{2.4}
\end{equation}
On ${\cal L}_{R}$ we can define the ''star product'' (i.e. the classical
operator related with the multiplication on $\widehat{{\cal L}_{R}}$): $symb(%
\widehat{f}\widehat{g})=symb\widehat{f}*symb\widehat{g}=(f*g)(\phi )$. It
can be proved (\cite{Wigner}, eq. (2.59)) that $(f*g)(\phi )=f(\phi )\exp
\left( -\frac{i\hbar }{2}\overleftarrow{\partial }_{a}\omega ^{ab}%
\overrightarrow{\partial }_{b}\right) g(\phi )=g(\phi )\exp \left( \frac{%
i\hbar }{2}\overleftarrow{\partial }_{a}\omega ^{ab}\overrightarrow{\partial 
}_{b}\right) f(\phi )$. We also define the Moyal bracket as the symbol
corresponding to the commutator in $\widehat{{\cal L}}_{R}$: $\{f,g\}_{mb}=%
\frac{1}{i\hbar }(f*g-g*f)=symb\left( \frac{1}{i\hbar }[f,g]\right) $. In
the limit $\hbar \rightarrow 0$ the star product becomes the ordinary
product and the Moyal bracket the Poisson bracket 
\begin{equation}
(f*g)(\phi )=f(\phi )g(\phi )+0(\hbar ),\qquad
\{f,g\}_{mb}=\{f,g\}_{pb}+0(\hbar ^{2})\ .  \label{2.11}
\end{equation}
Let us now consider the singular space $\widehat{{\cal L}}_{S,}$ the space
of the operators that commute with $\widehat{H}$ and $\widehat{P}$. We will
see that the mapping $symb$ given by (\ref{2.4}) is also well defined for
the observables in $\widehat{{\cal L}}_{S}$. In fact, from eq. (\ref{2.2})
we know that $\widehat{O}_{S}=\int_{p}\int_{0}^{\infty }O(\omega ,p)|\omega
,p\rangle \langle \omega ,p|d\omega dp^{N-1}$ so using well known procedures
we can conclude that $\widehat{O}_{S}=O(\widehat{H},\widehat{P)}$. But, if
the $\widehat{f}$, $\widehat{g}$ commute, as the members of the CSCO do, we
have $symb(\widehat{f},\widehat{g})=(f*g)(\phi )=f(\phi )g(\phi )+0(\hbar
^{2})$ then (using the same procedure as before) $symb\widehat{O}%
_{S}=O_{S}(\phi )=O(H(\phi ),P(\phi ))+0(\hbar ^{2})$. We have succeeded in
computing all the $symb$s of the observables of $\widehat{{\cal L}}_{S}$ up
to $0(\hbar ^{2}),$ so when $\hbar \rightarrow 0$ they are just $O(H(\phi
),P(\phi )),$ and we have defined the mapping $symb:\widehat{{\cal L}}%
_{S}\rightarrow {\cal L}_{S}$, $symb\widehat{O_{S}}=O_{S}(\phi )=O(H(\phi
),P(\phi ))+0(\hbar ^{2})$. Let us observe that if $O(\omega ,p)=\delta
(\omega -\omega ^{\prime })\delta (p-p^{\prime })$ we have from the last
equation 
\begin{equation}
symb|\omega ^{\prime },p^{\prime }\rangle \langle \omega ^{\prime
},p^{\prime }|=\delta (H(\phi )-\omega ^{\prime })\delta (P(\phi )-p^{\prime
})  \label{2.21}
\end{equation}
where we have disregarded the $0(\hbar ^{2})$ as we will always do below.
Therefore, from the eq. (\ref{2.4}) we have defined a classical space ${\cal %
A=L}_{R}\oplus {\cal L}_{S}$ and a mapping $symb:\widehat{{\cal A}}%
\rightarrow {\cal A}$, $symb\widehat{O}=O(\phi )$. Then we now have the
limit $\hbar \rightarrow 0$ for $\widehat{{\cal A}}\rightarrow {\cal A}$
(where also eq. (\ref{2.11}) is valid). But, even if this limit is well
defined and can be considered as {\it the classical limit of the algebra of
the operators,} it is only the limit of the {\it equations\/} of the system,
since these are a consequence of the algebra.

\section{The formalism for the states.}

Let us now consider the quantum and classical state sets and their limits.
Let us introduce the symbols $|\omega ,p)=|\omega ,p\rangle \langle \omega
,p|$ and $|\omega ,\omega ^{\prime },p,p^{\prime })=|\omega ,p\rangle
\langle \omega ^{\prime },p^{\prime }|.$ $\{|\omega ,p,p^{\prime })\}$ is
the basis of $\widehat{{\cal L}}_{S}$ and $\{|\omega ,\omega ^{\prime
},p,p^{\prime })\}$ is the basis of $\widehat{{\cal L}}_{R}$, then eq. (\ref
{2.2}) reads 
\begin{eqnarray*}
\widehat{O} &=&\int_{p}\int_{0}^{\infty }O(\omega ,p)|\omega ,p)d\omega
dp^{N-1} \\
&+&\int_{p}\int_{p^{\prime }}\int_{0}^{\infty }\int_{0}^{\infty }O(\omega
,\omega ^{\prime },p,p^{\prime })|\omega ,\omega ^{\prime },p,p^{\prime
})d\omega d\omega ^{\prime }dp^{N-1}dp^{\prime N-1}\ .
\end{eqnarray*}
The state are functionals over the space $\widehat{{\cal A}}=\widehat{{\cal L%
}}_{S}\oplus \widehat{{\cal L}}_{R}$. Therefore, let us consider the dual
space $\widehat{{\cal A}^{\prime }}=\widehat{{\cal L}_{S}^{\prime }}\oplus 
\widehat{{\cal L}^{\prime }}_{R}$. We will call $\{(\omega ,p|\}$ the basis
of $\widehat{{\cal L}_{S}^{\prime }}$ and $\{(\omega ,\omega ^{\prime
},p,p^{\prime }|\}$ the basis of $\widehat{{\cal L}_{R}^{\prime }}$. Observe
that $(\omega ,\omega ^{\prime },p,p^{\prime }|=|\omega ,p\rangle \langle
\omega ^{\prime },p^{\prime }|$ but $(\omega ,p|\neq |\omega ,p\rangle
\langle \omega ,p|$. Moreover 
\[
(\omega ,p|\omega ^{\prime },p^{\prime })=\delta (\omega -\omega ^{\prime
})\delta ^{N-1}(p-p^{\prime }) 
\]
\begin{equation}
(\omega ,\sigma ,p,s|\omega ^{\prime },\sigma ^{\prime },p^{\prime
},s^{\prime })=\delta (\omega -\omega ^{\prime })\delta (\sigma -\sigma
^{\prime })\delta ^{N-1}(p-p^{\prime })\delta ^{N-1}(s-s^{\prime })
\label{2.7}
\end{equation}
and the rest of the $(.|.)=0$. Then a generic functional of $\widehat{{\cal A%
}^{\prime }}$ reads 
\[
\widehat{\rho }=\int_{p}\int_{0}^{\infty }\rho (\omega ,p,)(\omega
,p|d\omega dp 
\]
\begin{equation}
+\ \int_{p}\int_{p^{\prime }}\int_{0}^{\infty }\int_{0}^{\infty }\rho
(\omega ,\omega ^{\prime },p,p^{\prime })(\omega ,\omega ^{\prime
},p,p^{\prime }|d\omega d\omega ^{\prime }dpdp^{\prime }\   \label{2.8}
\end{equation}
As functions $O(\omega ,p)$, $O(\omega ,\omega ^{\prime },p,p^{\prime })$
functions $\rho (\omega ,p)$, $\rho (\omega ,\omega ^{\prime },p,p^{\prime
}) $ are regular and have all the mathematical properties to make the
formalism successful \cite{Pra}. Moreover the $\widehat{\rho }$ must be
self-adjoint, and its diagonal $\rho (\omega ,p)$ must represent
probabilities, thus 
\begin{equation}
\widehat{\rho }=\widehat{\rho }^{\dagger }\ ,\qquad \qquad
\int_{p}\int_{0}^{\infty }\rho (\omega ,p)d\omega dp^{N-1}=1\ ,\qquad \qquad
\rho (\omega ,p)\geq 0\ .  \label{3.5}
\end{equation}
The $\widehat{\rho }$ with these properties belongs to a convex set $%
\widehat{{\cal S}},$ the set of the states. Also $(\widehat{\rho }|\widehat{O%
})=\int_{p}\int_{0}^{\infty }\rho (\omega ,p)O(\omega ,p)d\omega
dp^{N-1}+\int_{p}\int_{p^{\prime }}\int_{0}^{\infty }\int_{0}^{\infty }\rho
(\omega ,\omega ^{\prime },p,p^{\prime })O(\omega ^{\prime },\omega
,p^{\prime },p)$ $d\omega d\omega ^{\prime }dp^{N-1}dp^{\prime N-1}$. As $%
\widehat{{\cal L}}_{R}$ is a space of operators on a Hilbert space ${\cal H}$
so it is equal to its dual $\widehat{{\cal L}_{R}^{\prime }}$. Then it is
known that the symbol for any $\widehat{\rho }_{R}\in \widehat{{\cal L}%
_{R}^{\prime }}$ is defined as $\rho _{R}(\phi )=(\pi \hbar )^{-N}symb%
\widehat{\rho }_{R}$ by the same equations (\ref{2.4}). From this definition
we have (cf. \cite{Wigner} eq. (2.13)) 
\begin{equation}
(\widehat{\rho }_{R}|\widehat{O}_{R})=(symb\widehat{\rho }_{R}|symb\widehat{%
O_{R}})=\int d\phi ^{2N}\rho _{R}(\phi )O_{R}(\phi )  \label{3.8}
\end{equation}
and in $\widehat{{\cal L}}_{R}$ and $\widehat{{\cal L}_{R}^{\prime }}$ all
the equations are the usual ones (i. e. those of paper \cite{Wigner}).

Let us now consider the singular dual space $\widehat{{\cal L}_{S}^{\prime }}
$. In this space we will {\it define} the $symb\widehat{\rho }_{S}$ as the
function on ${\cal M}$ that satisfies a equation similar to (\ref{3.8}) for
any $\widehat{O}_{S}\in \widehat{{\cal L}}_{S},$ namely 
\begin{equation}
(symb\widehat{\rho }_{S}|symb\widehat{O_{S}})\circeq (\widehat{\rho }_{S}|%
\widehat{O}_{S})  \label{3.9}
\end{equation}
.We must define the meaning of these symbols, precisely $|symb\widehat{O_{S}}%
)$ and the inner product we are using in $(symb\widehat{\rho }_{S}|symb%
\widehat{O_{S}})$. From eq. (\ref{2.21}) we know that $symb|\omega ^{\prime
},p^{\prime })=\delta (H(\phi )-\omega ^{\prime })\delta ^{N-1}(P(\phi
)-p^{\prime })$. It is clear that we cannot normalize this function with the
variables and in the domain of integration of (\ref{3.8}). In fact using the
canonical variables $\alpha ,$ conjugated to the $P$, and $-t$, the
canonical variable conjugated to $H,$ for any function like $symb|\omega
^{\prime },p^{\prime })=\delta (H(\phi )-\omega ^{\prime })\delta
^{N-1}(P(\phi )-p^{\prime })$, which is a constant for the $\alpha $ and the 
$t$, the integral will turn out to be infinity. So these functions $f(H,P)$
are not classical densities since they do not belong to $L_{1}$ and they
must be normalized in a different way. But let us observe that $O_{S}(\phi
)=O(H(\phi ),P(\phi ))$ also is a function of this class. Then we can
normalize these functions if:

{\it i)\/} We only integrate over the momentum space $H,P$, precisely $%
||O_{S}(\phi )||=\int dH\int dP^{N-1}\int_{p}\int_{0}^{\infty }|O(\omega
,p)|\delta (H-\omega ^{\prime })\delta ^{N-1}(P-p^{\prime })dp^{N-1}d\omega
=\int dH\int dP^{N-1}|O(H,P)|$.

{\it ii)\/} We choose the regular function $O(\omega ,p)$ in the space $%
L_{1} $ of the momentum space $\omega ,p$, precisely $\int d\omega \int
dp^{N-1}|O(\omega ,p)|<\infty $. So we will normalize the $f(H,P)$ in this
way and we will perform {\it all the integrations in the\/} ${\cal L}_{S}$ 
{\it space\/} in the same way. Now we can calculate the function $\rho
_{S\omega p}(\phi )=symb(\omega ,p|$ which from eqs. (\ref{2.21}) and (\ref
{3.9})must satisfy 
\begin{equation}
\int dH\int dP^{N-1}\rho _{S\omega p}(\phi )\delta (H-\omega ^{\prime
})\delta ^{N-1}(P-p^{\prime })=\delta (\omega -\omega ^{\prime })\delta
(p-p^{\prime })  \label{3.14}
\end{equation}
where we have used the just defined way of integration. Then $\rho _{S\omega
^{\prime }p^{\prime }}(\phi )=symb(\omega ^{\prime },p^{\prime }|=\delta
\left( H(\phi )-\omega ^{\prime }\right) \delta ^{(N-1)}\left( P(\phi
)-p^{\prime }\right) $. So finally we can say that $(\widehat{\rho }_{S}|%
\widehat{O}_{S})\circeq (symb\widehat{\rho }_{S}|symb\widehat{O_{S}})=\int
dHdP^{n-1}\rho _{S}(\phi )O_{S}(\phi )$ the analogous of eq. (\ref{3.8}) in
space ${\cal L}_{S}$. Moreover 
\begin{eqnarray*}
\rho _{S}(\phi ) &=&\int_{p}\int_{0}^{\infty }\rho (\omega ,p)symb(\omega
,p|d\omega dp \\
&=&\int_{p}\int_{0}^{\infty }\rho (\omega ,p)\delta \left( H(\phi )-\omega
\right) \delta ^{(N-1)}\left( P(\phi )-p\right) d\omega dp=\rho (H(\phi
),P(\phi ))
\end{eqnarray*}
and it is a constant of the motion. It can be normalized in space ${\cal L}%
_{S}$ as $O(\omega ,p)$ namely 
\begin{equation}
\int d\omega \int dp^{N-1}|\rho (\omega ,p)|<\infty \qquad or\text{ }%
simply\qquad \int d\omega \int dp^{N-1}\rho (\omega ,p)=1  \label{3.200}
\end{equation}
if we take into account eq. (\ref{3.5}$_{2}$). Thus, according to the
prescriptions above, we have defined the mapping of the quantum states space 
$\widehat{{\cal A}^{\prime }}$ on the ``classical'' state space ${\cal A}%
^{\prime }$: $symb:\widehat{{\cal A}^{\prime }}\rightarrow {\cal A}^{\prime
} $ In the limit $\hbar \rightarrow 0$ equation (\ref{2.11}) is always valid
so we would have something like the classical limit for the states. But {\it %
it is not so}, because in general the obtained $\rho (\phi )$ {\it does not
satisfy} the condition $\rho (\phi )\geq 0$ (for every $\phi \in {}$) even
if $\hbar \rightarrow 0$. So $\rho (\phi )$ is not a density function and
therefore mapping (\ref{2.4}) is not a mapping of quantum mechanics on
classical statistical mechanics. Application (\ref{2.4}) does not lead to
the classical world. So for $\hbar \rightarrow 0$ the isomorphism (\ref{2.4}%
) is a mapping of quantum mechanics on a certain quantum mechanics ``alla
classica'', namely formulated in phase space ${\cal M}$ (but not with $\rho $
non-negative defined). Thus $\hbar \rightarrow 0$ is not the classical
limit. To obtain this limit we must introduce decoherence.

\section{Time evolution and the decoherence}

We only obtain the true classical limit for $t\rightarrow \infty $ via the
decoherence process. In fact, $\rho (t)$ evolves in time as $\rho (\phi
,t)=\rho (H(\phi ),P(\phi ))+\int_{p}\int_{p^{\prime }}\int_{0}^{\infty
}\int_{0}^{\infty }\rho (\omega ,\omega ^{\prime },p,p^{\prime })$ $%
e^{i(\omega -\omega ^{\prime })t/\hbar }symb(\omega ,\omega ^{\prime
},p,p^{\prime }|d\omega d\omega ^{\prime }dp^{N-1}dp^{\prime N-i}$. This
functional acts on spaces ${\cal A}$ giving 
\begin{eqnarray}
(\rho (\phi ,t)|O(\phi )) &=&\int dHdP^{N-1}\rho (H(\phi ),P(\phi ))O(H(\phi
),P(\phi )) \\
&+&\int_{p}\int_{p^{\prime }}\int_{0}^{\infty }\int_{0}^{\infty }\rho
(\omega ^{\prime },\omega ,p^{\prime },p)e^{i(\omega -\omega ^{\prime
})t/\hbar }O(\omega ,\omega ^{\prime },p,p^{\prime })  \nonumber \\
&&(symb(\omega ^{\prime },\omega ,p^{\prime },p|symb|\omega ,\omega ^{\prime
},p,p^{\prime }))d\omega d\omega ^{\prime }dp^{N-1}dp^{\prime N-1}\ . 
\nonumber  \label{3.222}
\end{eqnarray}
Now if the regular functions $\rho (\omega ,\omega ^{\prime },p,p^{\prime })$
and $O(\omega ^{\prime },\omega ,p^{\prime },p)$ are endowed with adequate
properties as those listed in \cite{Pra} the Riemann-Lebesgue theorem can be
used to obtain 
\begin{eqnarray*}
\lim_{t\rightarrow \infty }(\rho (\phi ,t)|O(\phi )) &=&\int dHdP^{N-1}\rho
\left( H(\phi ),P(\phi )\right) O(H(\phi ),P(\phi )) \\
&=&(\rho _{*}(\phi )|O(\phi ))
\end{eqnarray*}
for any $O(\phi )\in {\cal A}$ and where we have defined the functional $%
\rho _{*}(\phi )=\rho _{S}(\phi )$ and find the classical final limit. In
fact, we have found the weak limit $W$ $\lim_{t\rightarrow \infty }(\rho
(\phi ,t)|=(\rho _{*}(\phi )|$. As we can see only the singular part remains
and therefore, in the quantum case, only the singular-diagonal part remains
too, thus the time evolution has made the system decoheres. Of course, as
this kind of decoherence is obtained when $t\rightarrow \infty $ we can ask
ourselves which the decoherence time is. The problem is solved in papers\cite
{Pra} and \cite{Rolo}, where via an analytic continuation of the resolvent
of the von Neumann-Liouville operator in the complex plane, it is shown that
decoherence time is the inverse of the distance of the closer pole to the
real axis. Systems with no poles (e.g. the free particle) have infinite
decoherence time and therefore they do not decohere in practice.

Now we can find the property that was missing at the end of the last
section, i.e. $\rho (\phi )\geq 0$. In fact, condition (\ref{3.5}) yields $%
\rho _{*}(\phi )=\rho _{S}(\phi )=\rho \left( H(\phi ),P(\phi )\right) \geq
0 $ (cf. eq.(12)). So now we have the true statistical-classical limit when $%
t\rightarrow \infty ,\hbar \rightarrow 0,$ in which case the mapping $symb:%
\widehat{{\cal A}^{\prime }}\rightarrow {\cal A}^{\prime }$ maps the quantum
states with non-negative probability in the diagonal ($\rho (\omega ,p)\geq
0)$ into final classical states with $\rho _{*}(\phi )\geq 0,$ which are
non-negative density functions as they should be.

Moreover the $\rho _{*}(\phi )=\rho _{S}(\phi )$ of eq. (12) has a clear
physical meaning: It is the sum of densities strongly peaked in the
classical trajectories defined by the constant of the motion $H(\phi
)=\omega $, $P(\phi )=p$ averaged by the classical density function $\rho
(\omega ,p)$ which is properly normalized according eq. (\ref{3.200}). This
fact is essential in the localization process to obtain the final classical
limit (see \cite{Deco}).

In conclusion, we have demonstrated that ``symb'' and $\hbar \rightarrow 0$
give the correct classical-statistical limit of quantum mechanics for the
algebra of operators, but for the set of states it must be complemented with 
$t\rightarrow \infty ,$ which produces the decohered final state with a
non-negative density function.

\end{document}